\documentclass[
 aip,
 amsmath,
 amssymb,
 reprint,
 jcp,
 floatfix,
]{revtex4-2}

\usepackage[utf8]{inputenc}
\usepackage[T1]{fontenc}
\usepackage{mathptmx}
\usepackage{etoolbox}

\usepackage{graphicx}
\usepackage{dcolumn}
\usepackage{bm}
\usepackage{hyperref}

\usepackage[shortlabels]{enumitem}
\PassOptionsToPackage{backend=biber}{biblatex}

\usepackage{xcolor}
\definecolor{amethyst}{rgb}{0.6, 0.4, 0.8}

\usepackage{siunitx}

\graphicspath{ {DGraphics/} }

\begin{document}

\title{Cascading symmetry constraint during machine learning-enabled 
  structural search for sulfur induced Cu(111)-$(\sqrt{43}\times\sqrt{43})$ surface reconstruction}
\author{Florian Brix}

\author{Mads-Peter Verner Christiansen}
\author{Bjørk Hammer}
\email{hammer@phys.au.dk}
\affiliation{Center for Interstellar Catalysis, Department of Physics and Astronomy, Aarhus University, DK‐8000 Aarhus C, Denmark}

\begin{abstract}
  In this work, we investigate how exploiting
  symmetry when creating and modifying structural models may speed up
  global atomistic structure optimization. We propose a search
  strategy in which models start from high symmetry configurations and
  then gradually evolve into lower symmetry models. The algorithm is
  named \textit{cascading symmetry search} and is shown to be highly
  efficient for a number of known surface reconstructions.  We use our
  method for the sulfur induced Cu (111)
  $(\sqrt{43}\times\sqrt{43})$ surface reconstruction for which we
  identify a new highly stable structure which conforms with
  experimental evidence.
\end{abstract}

\maketitle

\section{Introduction}

The demand for new materials with specific properties to fuel
scientific and industrial advancement prompts an ever increasing
quest for materials prediction and discovery. Newly postulated
materials contribute to a wide range of applications, from
photocatalytic water splitting~\cite{Baker2022,PdSe2WaterSplitting} to
more effective battery
materials~\cite{D2CP05055B,RenwablesIssues,Wan2018}, new catalysts for
chemical synthesis of chemicals tackling the climate crisis, and a wide
range of electronic devices~\cite{Thermoelectric_PdSe2}.

Identification of the atomistic structure of the surfaces of solid
materials is often an important first step in characterizing the
physico-chemical properties of the materials. For inorganic materials,
the stability of a surface tends to be dictated by its total energy while
free energy terms are of minor importance. This means that atomistic
surface structure can be determined by means of global optimization in
combination with a reliable total energy expression, such as e.g.\
Density Functional Theory (DFT). Many such global optimization algorithms have been introduced ranging
from simple random structure search ~\cite{Pickard2011} and basin hopping methods ~\cite{wales1997global,wales1999global} to more
advanced methods such as minima hopping~\cite{goedecker2004minima}, simulated annealing~\cite{kirkpatrick1983optimization}, particle
swarm ~\cite{wang2012calypso}, evolutionary algorithms ~\cite{woodley1999prediction,woodley2004prediction,johnston2003evolving,oganov2006crystal,abraham2006periodic,wu2013adaptive,vilhelmsen2014genetic,paleico2020global,falls2020xtalopt,bauer2022systematic}, and novel strategies~\cite{pickard2019hyperspatial}.

In recent years, the catalog of global optimization methods has been
extended considerably via the introduction of machine learning techniques.
Various approaches have proven highly efficient in speeding up the
optimization. An often taken approach is to introduce a machine
learning interatomic potential (MLIP) with which the energy landscape may be probed
computationally cheaper than at the full DFT level. Here
the MLIP may be pretrained models based on DFT data known prior to the
global optimization~\cite{chen2018rise,ryan2018crystal,xu2019deep} or the
MLIPs may emerge from active learning protocols and be built on-the-fly
while the global optimization proceeds and accumulates data at the DFT
level~\cite{li2015molecular, tran2018active,jinnouchi2019fly,jennings2019genetic, zhang2020nwpesse,loeffler2020active,timmermann2021data, lourencco2021new}.

The MLIPs come in many forms with Gaussian Process Regression
~\cite{deringer2017machine, del2019local,hajibabaei2022fast,ronne2022atomistic}
and artificial neural
networks~\cite{behler2007generalized, ouyang2015global, schutt2018schnet,yang2021machine,han2021global,NEURIPS2022_4a36c3c5,batzner20223,wang2022accelerated,wanzebock2022neural},
being the the most often used means of modelling the total
energy. Other ways to use machine learning in global optimization
involve utilizing Bayesian statistics in selection of structural
candidates to be evaluated at the DFT
level~\cite{jorgensen2018exploration, todorovic2019bayesian, bisbo2020efficient,bisbo2022global, merte2022structure,wang2023magus}
and using image recognition and reinforcement learning ~\cite{putin2018reinforced, jorgensen2019atomistic, zhou2019optimization, mortensen2020atomistic, simm2020reinforcement,christiansen2020gaussian,meldgaard2020structure} and other generative models such as diffusion ~\cite{xie2021crystal,lyngby2022data} and generative adversarial networks (GANs)
to directly
construct the structural models~\cite{liu2023self}.

Despite the many advances in structural optimization techniques, it
remains a highly efficient strategy to exploit symmetry whenever
possible.

A common approach in work done so far is to generate
highly symmetric structures as first guesses ~\cite{wheeler2007sass,
  Pickard2011, wang2012calypso, lyakhov2013new, avery2017randspg,
  oganov2018crystal, zhao2023physics} or to bias the searches towards
highly symmetric structures~\cite{huber2023targeting} .

As an example, Shao \textit{et al}\cite{shao2022symmetry}
recently demonstrated how otherwise intractable structural problems
could be solved once the searches were limited to symmetric
structures, as specified by a space group. In general, exploiating
symmetry, problems involving many tens of atoms and hence readily hundreds 
of atomic coordinates may be mapped onto problems with only few
atoms and some tens of atomic coordinates. These problems are more
tractable as follows directly from the strong scaling of the size of
the configurational space with the number of atomistic degrees of
freedom.

In the present work, we have investigated the degree to which symmetry
may help to speed up structural search for a wide variety of surface
reconstructions. We conduct the searches in two ways: 1) Either the
symmetry is considered known and kept fixed all along the search, or
2) the symmetry is considered unknown and is changed dynamically
throughout the search in a cascading manner. In the latter case, the symmetry starts from
being the highest possible allowed by the surface unit cell and is gradually
lowered until it ends up being absent. We find that imposing a known
symmetry during a search is always advantageous compared to an unsymmetrized
search. We further find that the cascading approach is often as
efficient as using a fixed symmetry, which we attribute to highly
symmetric structures representing good seeds for structures with the
correct lower symmetry.

The paper is outlined as follows: First we describe how the symmetry
is handled during the structural searches. This involves protocols for
making completely new structures and for modifying existing
structures. We proceed by inspecting the evolution of structural
searches with various settings for the symmetry. Next, we demonstrate
how a number of known surface structures can be recovered by both the
fixed-symmetry and cascading-symmetry search methods. Finally, we
address a hitherto unsolved problem, the sulfidized
Cu(111)-$(\sqrt{43}\times\sqrt{43})$, and use the
cascading-symmetry method to identify a Cu$_{12}$S$_{12}$ overlayer structure
responsible for this surface reconstruction.

\section{Method}
Structural optimization can proceed according to a wide range of
algorithms. The simplest such are the random structure search (RSS) and
basin hopping (BH) methods, which both rely on the construction of a structural
candidate and subsequent relaxation of the candidate. In RSS, the
structural candidate is always constructed from scratch, while this is
only the case in the first iteration of a BH search. In subsequent BH
search iterations, the new candidate is constructed by modification of
a \textit{seed} candidate. The BH search algorithm further includes a selection
step in which it is decided if the newly constructed candidate is to become
the future seed candidate. More advanced search algorithms (see end of this
section) share the property with RSS and BH that a means of creating
structures either from scratch or via modification of known structures is
required. We therefore start by presenting how that can be done while constraining 
the structures to a chosen symmetry group. 

\subsection{Symmetry of surface structures}
For a periodic bulk material, the translational symmetry is
characterized by one of 14 different Bravais lattices. Considering the
atomic positions in the basis, and combining translations, rotations,
reflections, and inversions, it further turns out that periodic bulk
materials belong to one of 230 different space groups. Once a specific
surface is cut for a bulk crystal, the symmetry is reduced and only
five different lattice types and 17 different so-called
\textit{wallpaper} symmetry groups remain possible.

\begin{figure}[ht!]
\begin{center}
\includegraphics[width=0.5\textwidth]{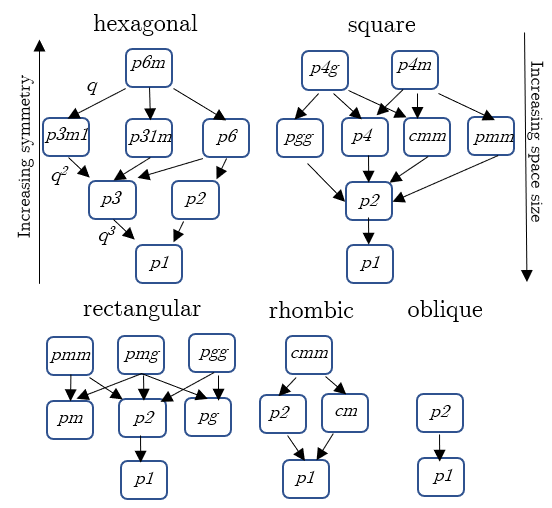}
\end{center}
\caption{\label{Fig_strat} The five different lattice types that exist
  for crystal surfaces and the corresponding allowed wallpaper
  groups. The symmetry groups of with smallest irreducible wedges are shown first, and
are connected with arrows to subgroups of lower symmetry and larger
irreducible wedge. The notation
from crystallography of bulk materials is used. The $q$-annotation
indicates the likelihood of reducing the symmetry per candidate
construction cycle when the cascading strategy is applied.}

\end{figure}

\begin{figure}[ht!]
\begin{center}
\includegraphics[width=0.5\textwidth]{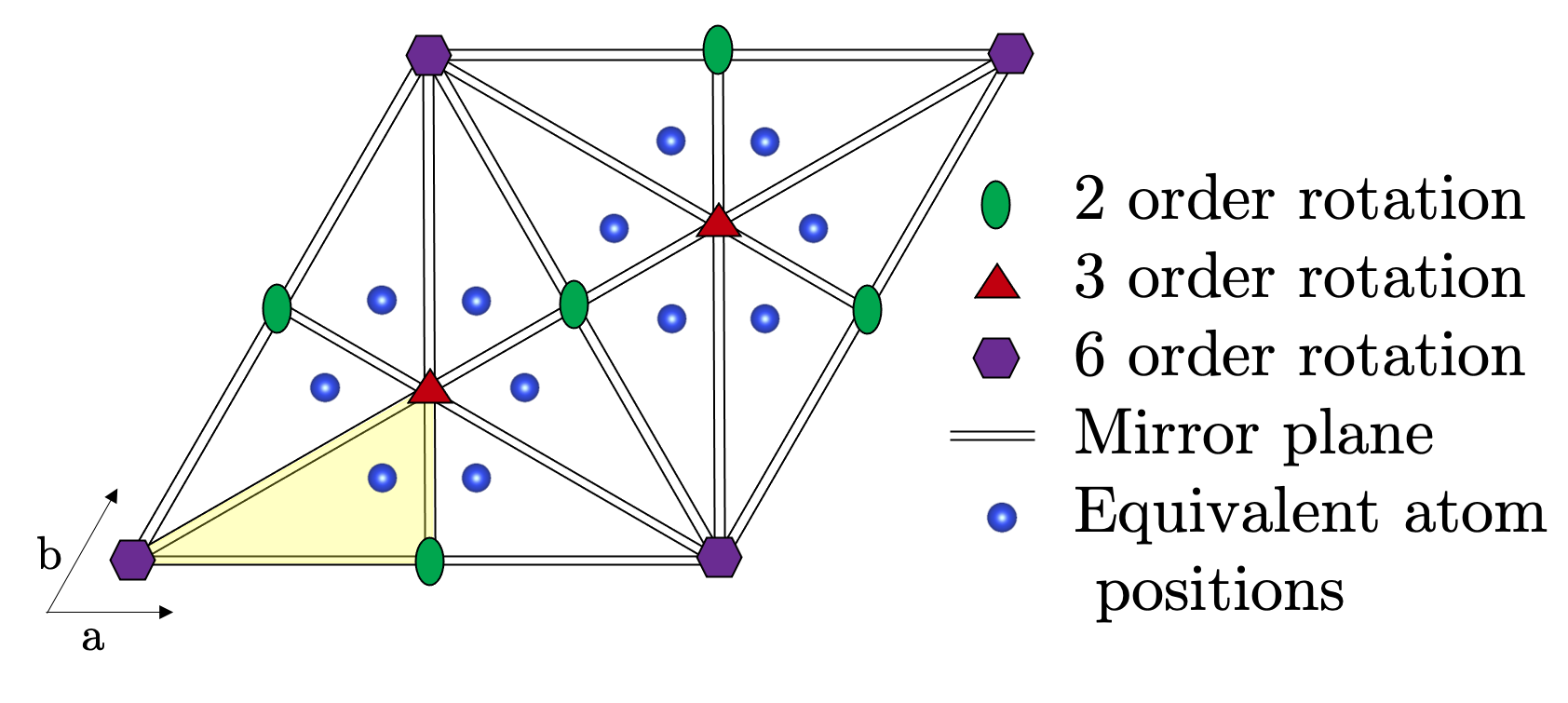}
\caption{\label{pmm} Rotational centers and reflection planes for a
hexagonal surface unit cell having the $p6m$ wallpaper symmetry
group. There are three 2-fold rotational centers, two 3-fold centers, and one
6-fold center, some of which are depicted multiple times
due to the periodicity. In addition, there are six reflection planes, which
may also appear several times due to the periodic condition. The yellow color
highlights the irreducible wedge which is where atoms can be placed
independently. Once placed in the irreducible wedge, the positions of
equivalent atoms in the rest of the surface cell follows from the
symmetry operations. This is illustrated for one atom (the blue dot).
}
\end{center}
\end{figure}

The five lattice types, hexagonal, square, rectangular, rhombic, and oblique,
follow from the shape of the surface unit cell, and the
wallpaper group, e.g.\ $p1$ or $p2$, is determined by the layer-wise
atomic positions of the material. Figure \ref{Fig_strat} gives an overview
of which wallpaper groups are possible for each lattice
type and Fig.\ \ref{pmm} depicts as an example the symmetry elements
of the $p6m$ wallpaper group of relevance to a hexagonal lattice type.
The irreducible wedge of the surface unit cell illustrated in yellow
color in Fig.\ \ref{pmm} represents the region in which atoms may be
placed independently while atomic positions in the rest of the unit
cell then follow from the symmetry operations of the wallpaper group.

In Fig.\ \ref{Fig_strat}, the size of the irreducible wedge has been
used to organize the wallpaper groups. For each lattice type, starting
from the top, the irreducible wedge is small and only some atoms can
be placed independently. Following the arrows to a symmetry subgroup,
symmetry operations are removed and the size of the irreducible wedge
increases until the $p1$ wallpaper group is reached, where the
irreducible wedge coincides with the entire surface unit cell and
every atom can be placed independently.  We stress that Fig.\
\ref{Fig_strat} only illustrates how wallpaper groups are related as
subgroups of \textit{increasing} irreducible wedge size, and that subgroup
relations for unchanged irreducible wedge size are omitted for clarity.

In the present work, we formulate symmetry-aware methods for
constructing structural candidates for crystal surfaces, including
surface reconstructions, and surface thin films. The surfaces will be
described by a set of atoms placed within a periodic surface unit cell
ontop of a periodic slab, i.e.\ a number of bulk layers. While the
lattice type is determined by the surface unit cell, a choice must be
made for the wallpaper symmetry group. We detail below how that is
decided upon in actual searches, but it is important to note at this
point, that we neglect any impact on the symmetry group from the atoms
in the slab -- only the atoms in the surface layer(s) are used when
deriving and discussing the wallpaper group of a structural
candidate. Since we have implemented rotational, reflection, and
inversion operations
assuming that the origin maps onto itself, it is desirable (when using our
code) to choose the registry of the atoms in the bulk layers (the slab) so that
the origin becomes a high-symmetry point, but this restriction could
be lifted if more general symmetry routines were written.

When a symmetric structure is built, placing an atom somewhere results
in having to place one equivalent atom in each irreducible wedge of the surface unit cell. One of these wedges is shown as the yellow-colored region in Fig.\ \ref{pmm}. If an atom is placed at e.g.\ the blue point within this wedge, in order to maintain the $p6m$ symmetry, a set of 11 atoms must be placed at the
symmetrically equivalent positions, that are indicated by the blue
points outside the colored wedge. Likewise, placing an atom at the
boundary of the wedge leads to a demand for placing atoms elsewhere in
the surface cell to maintain symmetry, albeit at fewer places. The
total number of symmetry equivalent points in the entire cell for any
given point in the irreducible wedge will in the following be referred
to as the \textit{multiplicity} of the point under that symmetry. Note
that multiplicity is not to be confused with \textit{order} of
a high-symmetry point, which rather describes how many times a point maps
onto itself under the symmetry operations.

\subsection{Building from scratch}
The construction of a structural candidate from scratch involves
placing atoms in a computational cell. The cell may already contain
preplaced atoms, \textit{a template}, and it must be specified what
type and amount of atoms, \textit{the stoichiometry}, should be present in the
final structure. On top of this, a required symmetry, \textit{the
  wallpaper group}, of the structure may be defined. If no symmetry
is defined, a random high-symmetry wallpaper group is selected among the ones
with the smallest irreducible wedge
for the defined surface unit cell (cf.\ Fig.\ \ref{Fig_strat}).
To fulfill the purpose of building such a structural candidate we
formulate a \textit{build-from-scratch} algorithm:
\begin{enumerate}
\item Pick a random atom type $Z$.
\item Evaluate how many atoms, $M_Z$, of type $Z$ are still to be placed in the cell.
\item Pick a random position, $\mathbf{x}$, whose multiplicity,
  $N(\mathbf{x})$, obeys $N(\mathbf{x})\le M_Z$.
\item Place atoms of type $Z$ at the $N(\mathbf{x})$ equivalent sites in the entire surface cell.
\item Loop until all atoms have been placed.
\end{enumerate}

\subsection{Building from a previous structure}
Once created, structures can be modified while respecting a given
wallpaper symmetry group (or subgroup) via a \textit{symmetric rattling} procedure. To rattle a
structure without changing the wallpaper group we introduce the
following algorithm:
\begin{enumerate}
\item Pick a random atom with index $i$.
\item Identify the multiplicity, $N(\mathbf{x}_i)$, of the position,
  $\mathbf{x}_i$, occupied by the atom.
\item Displace the atom randomly in the subspace of positions with
  multiplicity $N(\mathbf{x}_i)$. Move the symmetry-equivalent atoms accordingly.
\item Check distances between the displaced atoms. Wherever the
  distance is below a criterion: Merge the atoms.
\item If atoms were merged: Follow the build-from-scratch algorithm
  until the structure attains the original stoichiometry.
\item Loop for $P$ iterations.
\item Remove a random set of same-type atoms at low-multiplicity positions (i.e.\
  at high-symmetry positions).
\item Follow the build-from-scratch algorithm
  until the structure attains the original stoichiometry.
\item Loop for $Q$ iterations.
\end{enumerate}

The random positions specified in the algorithm have been implemented
as uniform displacements of the chosen atoms (within the allowed
subspace) with an amplitude from zero to some maximum distance.

\begin{figure}
\begin{center}
\includegraphics[width=0.5\textwidth]{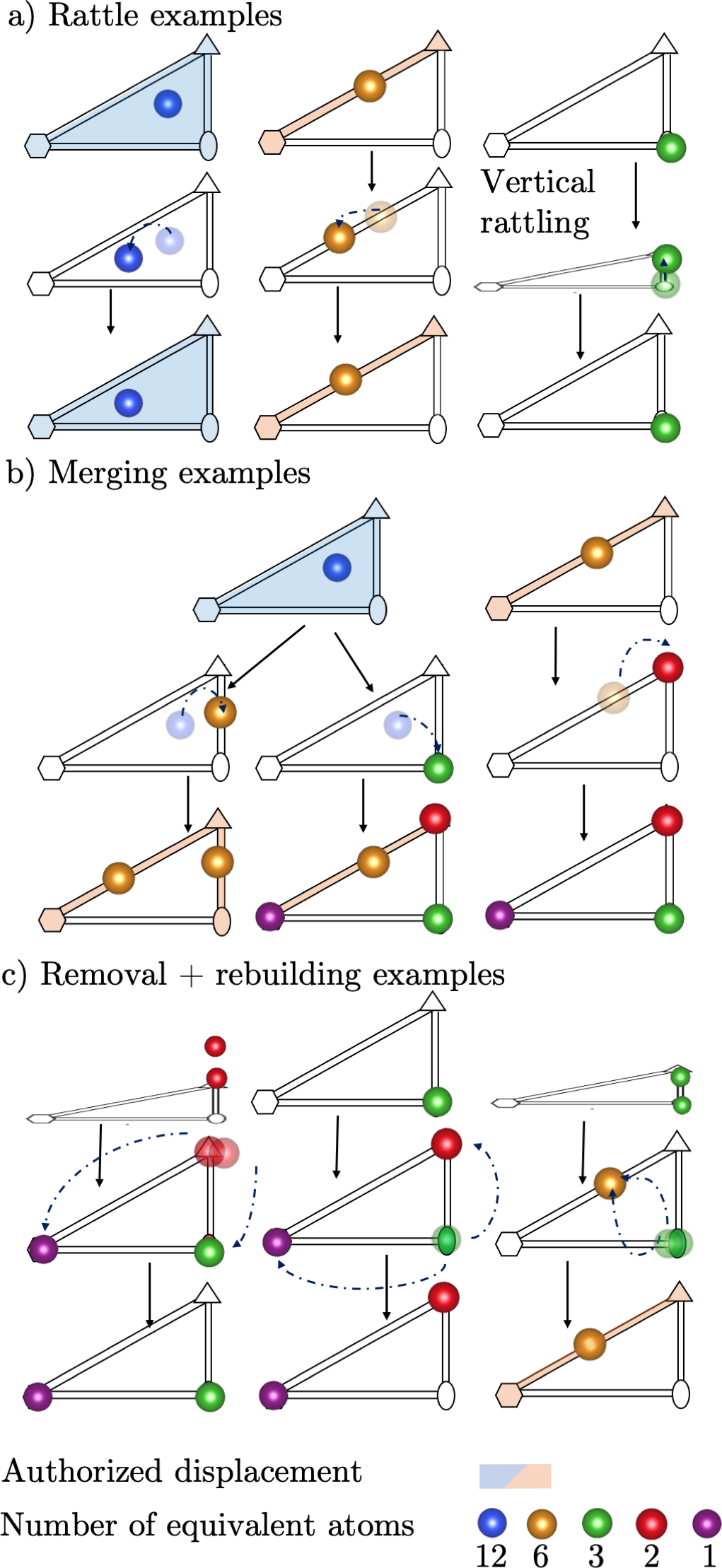}
\caption{Examples of the symmetry rattling algorithm. In three columns of 
(a) atoms of different multiplicity are displaced in their allowed subspace, 
without coming close enough to a mirror plane or high symmetry point that 
a merge is necessary. In (b) the displacement moves the atom to a mirror plane 
or high symmetry point and a merge that takes into account the multiplicity 
is required to keep the symmetry. Finally in (c) the removal and rebuilding 
process is exemplified, where atoms are removed and replaced in a way that 
keeps the total multiplicity constant, e.g. two atoms of multiplicity two 
are replaced with one atom of multiplicity three and one atom of multiplicity
one.}
\label{Fig1}
\end{center}
\end{figure}

Most elements of the algorithm are illustrated schematically in Fig.~\ref{Fig1}
for the irreducible wedge of the $p6m$ wallpaper group. In the figure, the colors are used to
show the multiplicity of positions, i.e.\ number of equivalent positions in the
entire unit cell. All atoms shown are of the same type, as the
algorithm only deals with manipulating same-type
atoms.

Figure \ref{Fig1}a illustrates the rattling of an atom in the
subspace of positions with a given multiplicity. After the rattling,
the atom and its replicas (not shown) are sufficiently far that no
merging of atoms is made.

Figure \ref{Fig1}b provides examples of rattling, that results in the
need for merging atoms. To the left, the blue atom (multiplicity of
12) is rattled to a position near a mirror plane and is merged with
an equivalent atom arriving to the same mirror plane in an adjacent
wedge. The same merging appears for other replica atoms and eventually,
12 atoms have turned into 6 atoms. Step 5 in the algorithm subsequently
assures that 6 new atoms be introduced in some way, here illustrated as
the appearance of an orange atom (multiplicity 6). In the middle
situation, the blue atom is rattled and merged into a green atom
(multiplicity 3) and step 5 causes the introduction of an orange, a
red, and a purple atom (i.e.\ with multiplicities 6, 2, 1) whereby the
number of atoms is maintained. Finally, to the right in Fig.\
\ref{Fig1}b, an orange atom (multiplicity 6) is rattled and merged
into a red atom (multiplicity 2) and a green and a purple atom
(multiplicities of 3 and 1) are added in step 5 so as to restore the
original number of atoms.

Figure \ref{Fig1}c illustrates the removal and rebuilding steps,
i.e.\ steps 7
and 8 of the algorithm. To the left, two red atoms (multiplicity 2)
are removed and a green and a purple atom are introduced (multiplicity
3 and 1). In the middle, a green atom (multiplicity 3)
is removed and a red and a purple atom are introduced (multiplicity
2 and 1). Finally, to the right, two green atoms (multiplicity 3) are
removed and an orange atom (multiplicity 6) is introduced.

\subsection{Building with cascading symmetry}
In the above algorithms, the symmetry must be decided upon before
structures can be built from scratch and subsequently rattled under
that given symmetry. Obviously, if the wallpaper group is known from experimental measurements, or if the best possible model structure of
a given symmetry is sought, the search can be performed with that
wallpaper group fixed throughout the search. We call this strategy:
\textit{Fixed symmetry rattling}.

However, an interesting option remains, namely to let the wallpaper
group vary throughout the search. We propose to start from the
wallpaper group with the highest
possible symmetry for the chosen surface cell
and then progressively explore less symmetric wallpaper groups by
rattling the candidates imposing less and less symmetry
constraints (for instance, starting with a $p6m$ candidate and rattling
it with $p6$ symmetry). We term this strategy: \textit{Cascading
symmetry rattling}. Specifically, this strategy is implemented by adding a 
10th step to the previously described fixed symmetry rattling
\begin{enumerate}
  \setcounter{enumi}{9}
  \item Decrease the symmetry to a random subgroup  with a lower order of symmetry  of the present
  wallpaper group with likelihood $q^{n+1}$, where $n$ is 
  the rung of the symmetry group starting with $n = 1$ for the most 
  symmetric groups and increases by 1 in each lower rung. 
\end{enumerate}
Thus, each time a structure is rattled, there is a
probability of decreasing the symmetry by selecting a subgroup of the
current wallpaper group, i.e.\ a less symmetric
compatible wallpaper group than the current one. The allowed
transitions between wallpaper groups can be assessed from Fig.~\ref{Fig_strat}.
  

Since the number of local minima increases exponentially with degrees of
freedom~\cite{oganov2018crystal,oganov2019structure}, the number of structures to consider
increases exponentially as we start to consider lower symmetry
wallpaper groups. Hence, there is a need to spend more and more optimization
attempts for a given wallpaper group as the search progresses and the
symmetry is lowered via the cascading protocol. The last step in the
symmetric rattle algorithm assures this behavior by imposing a power
law-depending likelihood for transitioning to a lower-symmetry
wallpaper group for future rattle-based candidate constructions.
Specifically the power law chosen is: $q^{n+1}$, 
where $q$ is a probability and 
$n$ is the rung of the symmetry group of the candidate.  This strategy is illustrated for the hexagonal lattice type
in Fig.~\ref{Fig_strat}, where the probabilities are shown next to
some of the transition
arrows. The search thus starts with highly symmetric
structures and then progressively explores less symmetric structures
along the search.

\subsection{GOFEE}

In the above discussion, the need for algorithms for candidate generation either from
scratch or from previously derived candidates was argued based on the
random structure search (RSS) and basin hopping (BH) methods. In the
present work, we use a related search algorithm, the Global
Optimization with First-principles Energy Expressions method (GOFEE)~\cite{bisbo2020efficient,bisbo2022global}
which combines elements of Bayesian statistics, machine learning, and
evolutionary algorithms. The GOFEE algorithm implemented in the
Atomistic Global Optimization X (AGOX) package framework
\cite{christiansen2022atomistic} built on the Atomistic Simulation Environment~\cite{hjorth2017ase} (ASE).

The GOFEE method is an iterative search method, that in each iteration 
does the following:
\begin{enumerate}
\item Create or modify a large number of structure candidates.
\item Relax them in the lower-confidence bound (LCB),
  $F=E-\kappa\sigma$, of an on-the-fly learnt machine learning interatomic
  potential, where $E$ is the total energy expectation, $\sigma$ is
  the associated uncertainty, and $\kappa$ is a constant.
\item Select the most promising candidate according to the LCB.
\item Evaluate the selected candidate at the DFT level.
\item Update the machine learning interatomic potential with the new DFT data.
\end{enumerate}
This is repeated for a set number of iterations reflecting the total 
computational budget.

By using the \textit{symmetric build-from-scratch} and
\textit{symmetric rattle} algorithms introduced in this work for
the candidate creation step of the GOFEE algorithm we obtain a
symmetry-aware GOFEE method, which -- via the symmetry-lowering
element of the \textit{symmetric rattle} algorithm -- performs the
structural search while cascading from high to low symmetry wallpaper groups.
The two algorithms have been implemented as "generator" modules for
the AGOX package and are available via gitlab. See the code availability 
section \ref{sec:code_section} for details. 

\section{Method benchmark}
The two new symmetric methods presented in this work have been applied
to example problems to quantify the relevance of such an approach in a
global optimization context. The GOFEE method with a global gaussian
process regression method has been used together to select relevant
candidates to be evaluated with the GPAW DFT computation
code~\cite{enkovaara2010electronic,mortensen2005real} and the
Perdew-Burke Ernzerhof functional~\cite{perdew1996generalized}.

\subsection{Benchmark of searching with fixed symmetry}
It is well known that the lowest-energy structure for the Si(111)
surface is the 7$\times$7 Dimer-Adatom-Stacking fault (DAS) reconstruction, which involves 102
atoms~\cite{lander1963structures}. In order to test the influence of
searching with various wallpaper groups, we first considered the smaller,
more tractable Si(111) system with a $5\times 5$ surface unit
cell. For this system, the lowest-energy structure also attains a
DAS reconstruction, but involving only 50 
atoms. We note that Si(111) may indeed be observed to form this structure, when a Si
crystal is cleaved at room temperature and subsequently heated to
$350 ^{\circ}$C~\cite{feenstra1990formation}.

This surface reconstruction follows a $p6m$ wallpaper group which divides the cell into 12 irreducible wedges. The search was also conducted on lower symmetry wallpaper groups including $p6m$, $p6$, and $p3m1$ which divide the unit cell into 6 irreducible wedges, and $p3$ which divides the unit cell into 3.

\begin{figure*}
    \centering
    \includegraphics[width=1\linewidth]{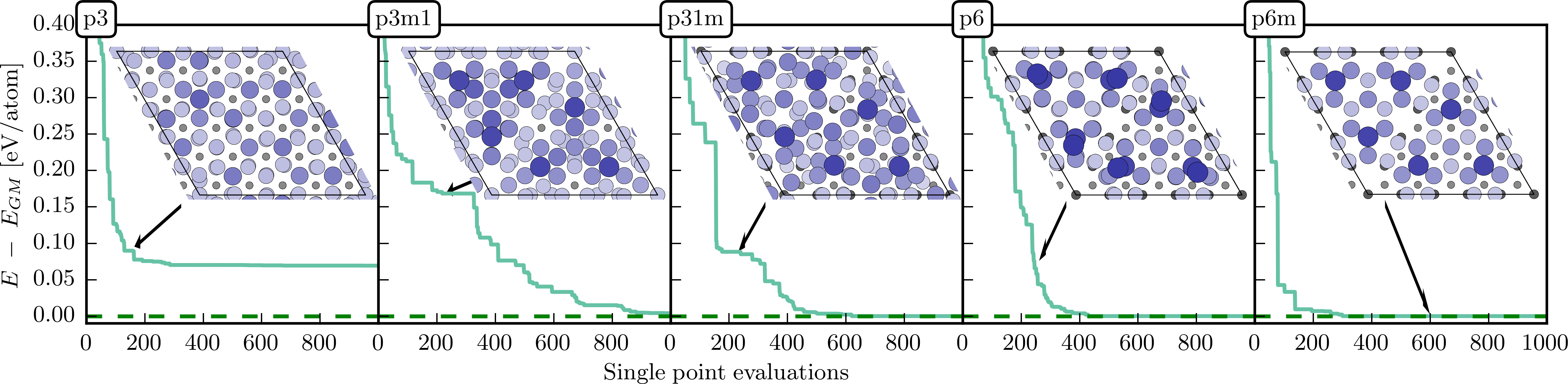}
    \caption{Energy curves obtained for selected individual searches
      with different wallpaper groups in the \textit{fixed symmetry
        rattling} approach. Energies are given relative to the
      lowest-energy $(5\times 5)$-DAS structure.}
    \label{fig:fixed_compar}
\end{figure*}

The searches for this surface reconstruction were done on two-layer
slabs of Si(111) in a 5$\times$5 surface unit cell. Dangling Si bonds
on the backside of the slab
were saturated by H. The GOFEE algorithm
was followed for 1000 iterations in each of which
60 symmetric candidates were created.
Figure \ref{fig:fixed_compar} shows five examples of the evolution of
the energy of the ever best structure during single search runs using
the fixed symmetry strategy. As the wallpaper groups used, $p3$, $p3m1$,
$p31m$, $p6$, and $p6m$, involve higher and higher symmetry, the
energy of the ever best structure quenches faster and faster being
within 100 meV/atom from the lowest-energy structure in 400 iterations
or less. Using all but the $p3$ wallpaper group, the final structure
at 1000 iterations is indeed the $5\times 5$ DAS structure.

\begin{figure}[ht!]
\begin{center}
\includegraphics[width=0.5\textwidth]{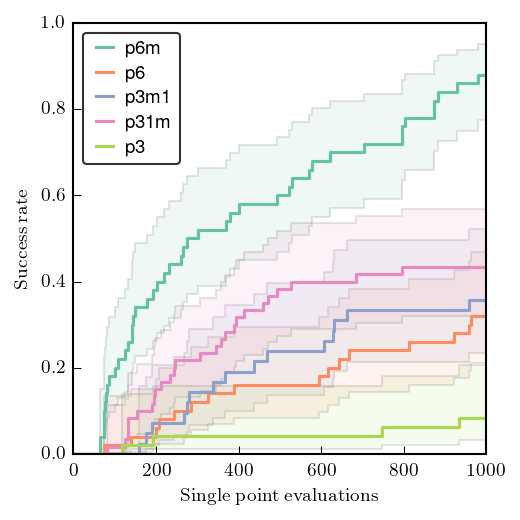}
\caption{Success curves using various fixed symmetry wallpaper groups
  while searching for the Si(111)-(5$\times$ 5) DAS
  reconstruction. Each curve is based on 50 independent search runs as
  the ones presented in Fig.\
  \protect\ref{fig:fixed_compar}. As an example, inspecting the
  success curve corresponding to using the $p3m1$ wallpaper group (the
  blue curve) e.g.\ reveals that about 20\% of searches have found the
  DAS structure in 400 iterations or less and that about 35\% have
  found it after 1000 iterations.}
\label{Fig2} 
\end{center}
\end{figure}

While Fig.\ \ref{fig:fixed_compar} presents individual search runs,
Fig.\ \ref{Fig2} compiles the results of a large number of
searches for each of the different fixed symmetries. The figure shows
\textit{success curves} that are computed as the the share of
independent runs that have found the DAS structure at or before a
given iteration. As an example, if 50 searches are conducted and 10
of them finds the solution within 400 iterations, the corresponding
success curve will pass through 20\% at 400 iterations. Thus, the
further to the left a success curve lies, and the closer it approaches
100\%, the better the underlying search method is.

Figure \ref{Fig2} clearly shows that as wallpaper groups of higher and
higher symmetry are used, the searches become more and more efficient.
In the $p3$ wallpaper group, the right structure is only found with a
probability lower than 10~\% in 1000 iterations whereas it increases
up to more than 85 \% in the most symmetric $p6m$ wallpaper
group. Leaving out the use of symmetry altogether, which corresponds
to using the $p1$ wallpaper group, the algorithm turns out incapable of finding the
$(5\times 5)$-DAS structure in 1000 iterations and is hence not shown in
the figure. Intermediate wallpaper groups, here $p3m1$, $p31m$ and
$p6$ are different equivalent ways to divide the hexagonal cell in
6. As expected, they all perform better than $p3$, and reach between 25 and 45 \%
in 1000 iterations.

\begin{figure}[ht!]
\begin{center}
\includegraphics[width=0.5\textwidth]{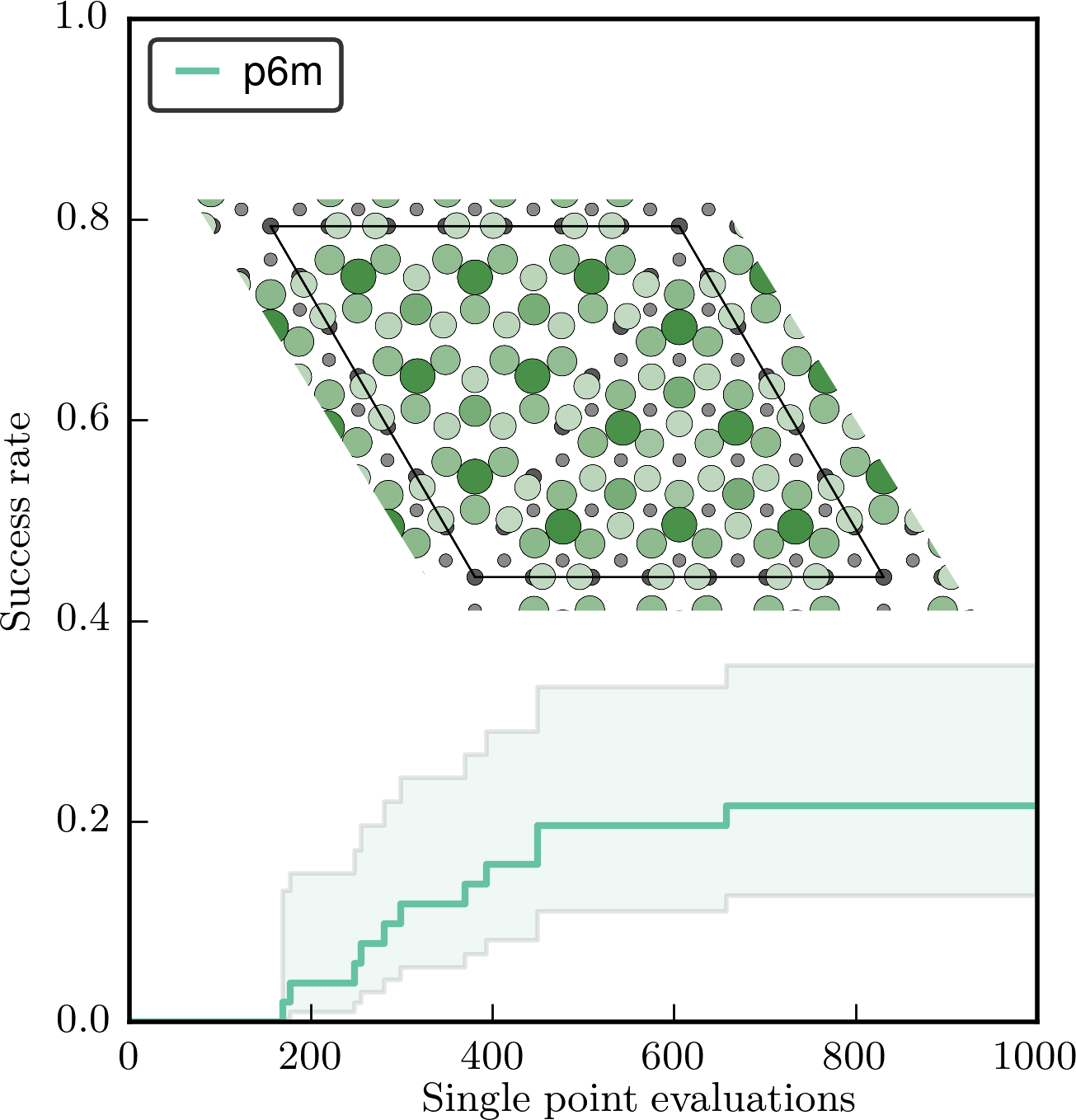}
\caption{Success curves using $p6m$ fixed symmetry wallpaper group
  while searching for the Si(111)-(7$\times$7) DAS
  reconstruction  based on 50 independent search runs.}
\label{Fig77} 
\end{center}
\end{figure}

Having seen that the Si(111)-$(5\times 5)$-DAS could be found
with a high success rate when utilizing the full $p6m$ wallpaper
group, we repeated the search for the larger Si(111)-$(7\times 7)$
system. Figure \ref{Fig77} presents the results of 50 individual
searches utilizing the $p6m$ wallpaper group. It is seen that the
Si(111)-$(7\times 7)$ DAS structure is found within 1000 
iteration in about 20\% of the cases which we consider to be a highly
satisfactory success rate.

\subsection{Benchmark of searching with cascading symmetry}

To investigate the cascading symmetry approach, we selected five
global optimization problems for complex surface reconstructions,
where in each case more than 12 atoms are involved and where the
solutions exhibit various symmetry wallpaper groups.

The first example is tin oxidation on Pt$_3$Sn (111) substrate showing
a $(4\times 4)$ reconstruction. The preferred stoichiometry and
structure were recently shown to be Sn$_{11}$O$_{12}$ exhibiting a
$p6$ wallpaper group~ \cite{merte2022structure}. It was modelled by
Sn$_{11}$O$_{12}$ placed in a $p(4\times 4)$ unit cell of Pt$_3$Sn
slab consisting of 2 fixed layers.

The second example is Ag$_2$O single layer oxide that forms in a
$p(4\times 4)$ cell on pure, close-packed
silver~\cite{schmid2006structure,schnadt2006revisiting,schnadt2009experimental}. This
surface reconstruction has recently been shown to be of interest for
the oxidation of ethylene compared to other surface reconstructions of
Ag oxide~\cite{jones2016oxidation}. The structure which was shown to
be stable is Ag$_{12}$O$_6$ forming a wallpaper group close to $p6m$
but distorted with O atoms not sitting at the same height of their
first neighbor so that the most stable structure exhibits a wallpaper
group of $p31m$.To study this structure, we deposited Ag$_{12}$O$_6$
on a 2 layers model slab of the $p(4\times 4)$
cell~\cite{schmid2006structure}.

The third example is the $3\times 3$ surface reconstruction on SiC(111).
Its structure was identified through a combination of
DFT calculations and experimental techniques in
1998~\cite{starke1998novel}. Starke et al.\ showed that its formation
mechanism involves a relaxation form the most symmetric SiC bulk into
a lower symmetry wallpaper group with only a 3 order rotation and
therefore a $p3$ wallpaper group. For this system, we deposited 13 Si atoms on 2 fixed layers
of the SiC(111) surface in a  $(3\times 3)$ cell.

The fourth example is VO$_3$ deposited on Rh(111)-$(\sqrt{13}\times
\sqrt{13})$ for which the structure was discovered by Schoiswohl et
al. \cite{schoiswohl2004atomic} which follows the highest possible
symmetry that is the $p6m$ wallpaper group and involving more atoms
than the other benchmark structure studied in this work. This system
was modeled by depositing V$_6$O$_{18}$ on 2 fixed layers of Rh(111)
in a $(\sqrt{13}\times \sqrt{13})$ cell.

The last example is the well-known GaAs (001) stable in a square $c(4\times 4)$ 
unit cell~\cite{falta1992structure,karmo2022reconstructions,penev2004atomic}. 
GaAs in a direct band gap semiconductor with a zinc blende structure used for 
various transistor types and also as a substrate for growing other semiconductors 
in the zinc blende structure such as GaN ~\cite{lin1993p}.
The As atoms tend to form As$_2$ dimers at the surface to avoid
dangling bonds. The $c(4\times 4)$ reconstruction is a well known
example of this effect exhibiting a $cmm$ wallpaper group. This
surface reconstruction was modelled by depositing Ga$_8$As$_{14}$ on 3
layers on GaAs zinc blende structure. Dangling bonds of the As atoms
at the back side of the slab were saturated using H atoms.

\begin{figure*}
\begin{center}
\includegraphics[width=1\textwidth]{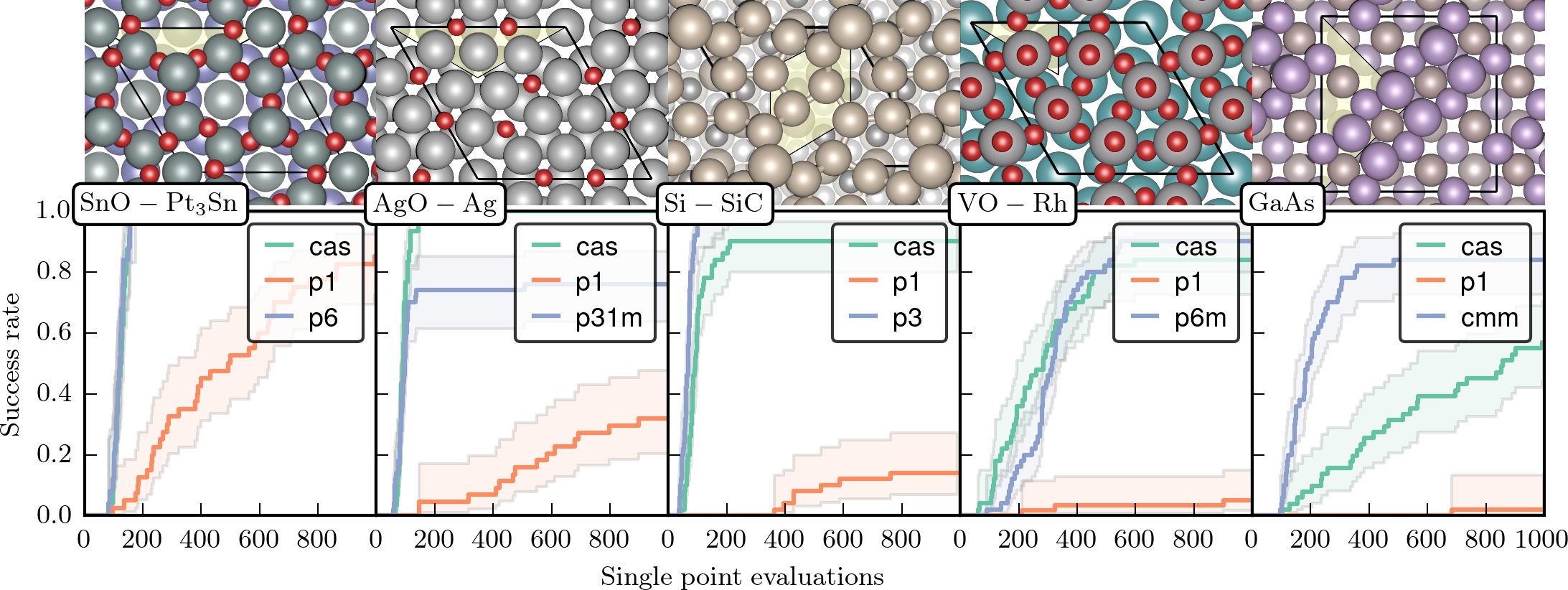}
\caption{\label{Fig3} Optimal structures and success curves for Pt$_3$Sn(111)-$p(4\times
  4)$-Sn$_{11}$O$_{12}$, Ag(111)-$p(4\times 4)$-Ag$_{12}$O$_6$,
SiC(111)-$p(3\times 3)$-Si$_{13}$, Rh(111)-$(\sqrt{13}\times                                                                    
\sqrt{13})$-V$_6$O$_{18}$, and GaAs(001)-$c(4\times
4)$-Ga$_8$As$_{14}$. Three different search strategies are used: The
cascading symmetry method (cas), no symmetry (p1), and the fixed
symmetry method ($p6$, $p31m$, $p3$, $p6m$, and $cmm$, respectively). In the
structure plots, the yellow shadings indicate the irreducible
wedges when using the fixed symmetry method.}
\end{center}
\end{figure*}

In Figure \ref{Fig3}, the performance of the cascading symmetry
approach is presented in the form of success curves for the five
different test systems. The likelihood of losing symmetry, $q$ is set
to 20\% for all searches. For each system, the method is capable of
finding the lowest-energy structure with a high likelihood in the 1000
GOFEE iterations that are done. In the figure, we also show the
success curves when solving the problems without use of symmetry
($p1$) or when using the fixed symmetry approach with the wallpaper
group that the solution has. Comparing the success curves reveals the 
cascading symmetry strategy always gives better results than when
omitting the use of symmetry, and that it in most cases performs as
well as when the correct symmetry is imposed in the fixed symmetry approach.

For the Sn oxidation on Pt$_3$Sn, the correct structure was found in every case in
less than 200 unique DFT evaluations, while 800 iterations were needed to reach 
the probability 80 \% without considering the symmetry. The symmetry search strategy 
shows comparable results to imposing the $p6$ wallpaper group.

For the second example, considering the symmetry reached roughly the same 
performance but the random search only reached 30 \% success in 1000 iterations. 
The symmetry search strategy performs better than imposing the $p31m$ wallpaper 
group which does not reach the solution in roughly 30 \% of independent searches.
 
For the three other systems, the right structures were found only in a few searches
within 1000 DFT calculations when symmetry was not used. For all of them, 
the right structures were found with a probability lower than 20 \%.
For SiC surface reconstruction, both the fixed symmetry strategy and cascading symmetry strategy
reached more than 85 \% success rate and was also increased to roughly 85 \% 
for the vanadium oxide on Rh (111). 

For all of these systems, this new strategy is much more efficient and can be 
expected to help finding structures that the symmetry unaware random generation 
and rattling could not reach in reasonable computing time for more complex systems,
involving more atoms or more local minima to explore.

\subsection{Discussion on the benefit of the symmetry cascade strategy}
For some of the examples, a few interesting points are worth clarifying. For the silver-oxide 
the search in the $p31m$ wallpaper groups levels off at 75 \%. 
This behavior can be attributed that fixing the $p31m$ wallpaper group for
this problem get stuck in a local minimum that the MLIP does not
overcome for some cases. The high variety of candidates produced by the cascading 
symmetry strategy allows it to escape these local minima and thus solving the 
problem with a higher rate of success. 

The SiC is one of the two case for which the cascading symmetry strategy performs 
slightly worse than fixing the right wallpaper group. This behavior is easily
explained by the fact that the symmetric strategy can explore the $p2$ wallpaper
group which is incompatible with the $p3$ wallpaper group of the solution, 
thus delaying finding the global minimum slightly. 

The GaAs surface reconstruction is the other case for which the cascading symmetry 
strategy does not perform as well as the fixed strategy. The explanation for this 
is the same as for the SiC system, but here more pronounced. For

square cells, the searches are started out in the two most symmetric
wallpaper groups, $p4m$ and $p4g$, and may in principle evolve into
structures of the correct $cmm$ wallpaper group. However, they have
three alternatives with similarly sized irreducible wedges (the $pmm$,
$pgg$, and $p4$, cf.\ Fig.\ \ref{Fig_strat}) meaning that the
cascading symmetry search ends up exploring more plane groups
incompatible with the $cmm$ solution than is the case in the hexagonal
examples, which rationalizes the decreased the success rate of the
cascading method for the GaAs system.

Nevertheless, the cascading symmetry strategy search gives good performances and satisfactory 
behaviors for all examples without imposing previous knowledge of the wallpaper group 
of the global minimum solution.
For all the tested cases, the method can find the lowest energy structure in more than 
60 \% of searches in less than 1000 DFT single-point calculations.
 
\section{The sulfur induced Cu (111) ( $\sqrt{43} \times \sqrt{43}$) surface reconstruction}

As further proof of the strength of this method we apply it to a sulfur-induced 
surface reconstruction of Cu(111), a system that been studied experimentally 
and is known to exhibit a $(\sqrt{43}\times\sqrt{43})$ unit 
cell \cite{wahlstrom1999observation,wahlstrom2001low}. Structural 
models have been proposed by Liu et al. with a Cu$_9$S$_{12}$ stoichiometry~\cite{liu2014search}, 
but without a definitive conclusion on the low-temperature structure observed. 


The copper content involved in this surface reconstruction is not known, 
but the sulfur content was estimated to be around 12 atoms per $(\sqrt{43}\times\sqrt{43})$
unit cell which corresponds to a surface coverage of 0.28 ML consistent with the 
fact that this structure is observed up to a coverage of 0.25 ML~\cite{wahlstrom1999observation}.
This structure is also described as a honeycomb-like structure, making it 
likely to present three-fold or six-fold rotations.

We chose to perform the search for a range of $N_{Cu}$:$N_{S}$ stoichiometries 
with $8 \le N_{Cu} \le 18$, $N_{S}$ = 11 or 12 sulfur atoms then compatible 
with the experimental observations. For each stoichiometry, 10 independent searches 
were carried out with a plane wave cutoff of 500 eV and 1 $k$-point.

Previous studies on CuS bulk and on CuS surfaces have shown that applying a DFT+U 
correction approach with U$_{\mathrm{eff}}$=5 eV was suitable to represent the localization of 
electrons over the Cu-S bond both for bulk study and surface properties~\cite{morales2014first,morales2017surfaces,paliwal2021fermi}. 
Therefore, we performed all calculations using the simplified rotationally invariant 
form of Dudarev et al. ~\cite{liechtenstein1995density,dudarev1998electron}. 
The structure optimization was done on a model slab including 2 layers of Cu(111) 
resulting in 86 Cu atoms for the substrate.

In order to take into account the relaxation of the substrate layers, 
structures within 1.5 eV of the lowest energy structure for each stoichiometry 
were subsequently relaxed on a thicker model slab of four layers with the two upper
layers free to relax and $2\times 2\times 1$ $k$-points. 
After which all the relaxed structures were compared to find the
energetically most favorable one.


The mean energy of formation per sulfur atom is calculated with respect to
the chemical potential of Cu, $\mu_{Cu}$ with the following formula:
\begin{equation}
 E_{f}= \frac{1}{N_{S}}\bigl[(E_{total}-E_{slab}) -N_{Cu} \mu_{Cu} \bigr] - \frac{1}{2}E_{S_{2,g}}
\label{eq:energy_gain}
\end{equation}
where E$_{S_{2,g}}$ is the energy of gaseous S$_2$ in a box and $\mu_{Cu}$ is 
the energy for adding one Cu atom on a kinked Cu (111) surface edge.
The Cu$_9$S$_{12}$ structure proposed by Liu et al. \cite{liu2014search} appeared
in the tested structures, but did not turn out to be among the most stable.

\begin{figure}[ht!]
\begin{center}
\includegraphics[width=0.45\textwidth]{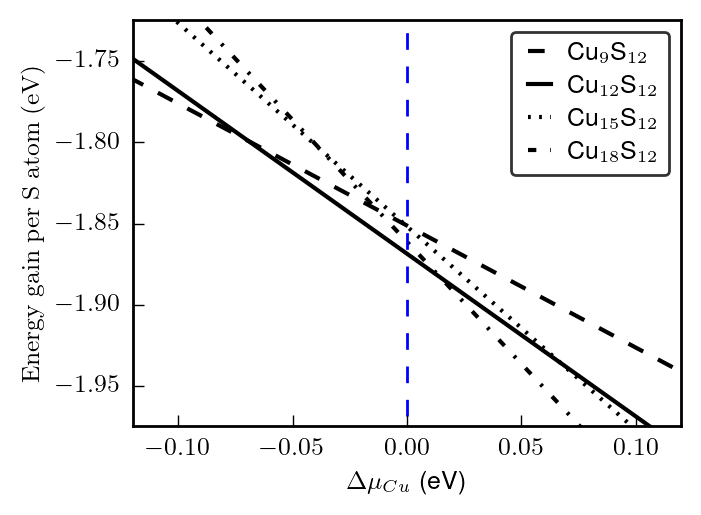}
\includegraphics[width=0.45\textwidth]{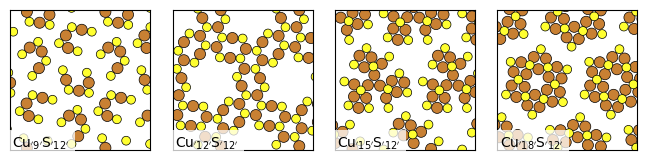}
\caption{Energetic comparison of the best structures for each stoichiometry (top) and the best 
structural models for the four most favorable stoichiometries (bottom).}
\label{Fig_struc_cus}
\end{center}
\end{figure}

Among the tested stoichiometries the best structures for Cu$_{8}$S$_{12}$,
Cu$_{12}$S$_{12}$ ,Cu$_{17}$S$_{12}$ Cu$_{18}$S$_{12}$ contained six-fold symmetry 
and for all other stoichiometries the best structures have at least a three-fold rotation.
Fig. \ref{Fig_struc_cus} shows a diagram of the stability, calculated with
Eq.\ \eqref{eq:energy_gain}.

In the figure, we introduce a term,
$\Delta \mu_{Cu}$, which represents a possible deviation of the
chemical potential of Cu from the calculated one (assuming any issues
with extracting it from super cells of different sizes than the one
used for the $(\sqrt{43}\times\sqrt{43})$ surface reconstruction).
In the figure, we only include structures with six-fold or three-fold symmetry, which require
the number of atoms to be a multiple of three or six, as these were
found to be the most favorable. We find the structure with 
Cu$_{12}$S$_{12}$ stoichiometry to be favored over a wide range of $\Delta\mu_{Cu}$
values around 0 eV meaning that we can safely assign this as the
computationally derived preferred structure. 
Two of the structures shown have stoichiometries being a multiple of
6, Cu$_{12}$S$_{12}$ and Cu$_{18}$S$_{12}$, which underlines the energetic gain obtained when 
forming a sixfold symmetric structure for this system. The two others present 
a three-fold symmetry. The bottom row of the Fig. \ref{Fig_struc_cus} shows illustrations
of four structures.

\begin{figure}
\begin{center}
\includegraphics[width=0.3\textwidth,trim={600 300 600 250},clip]{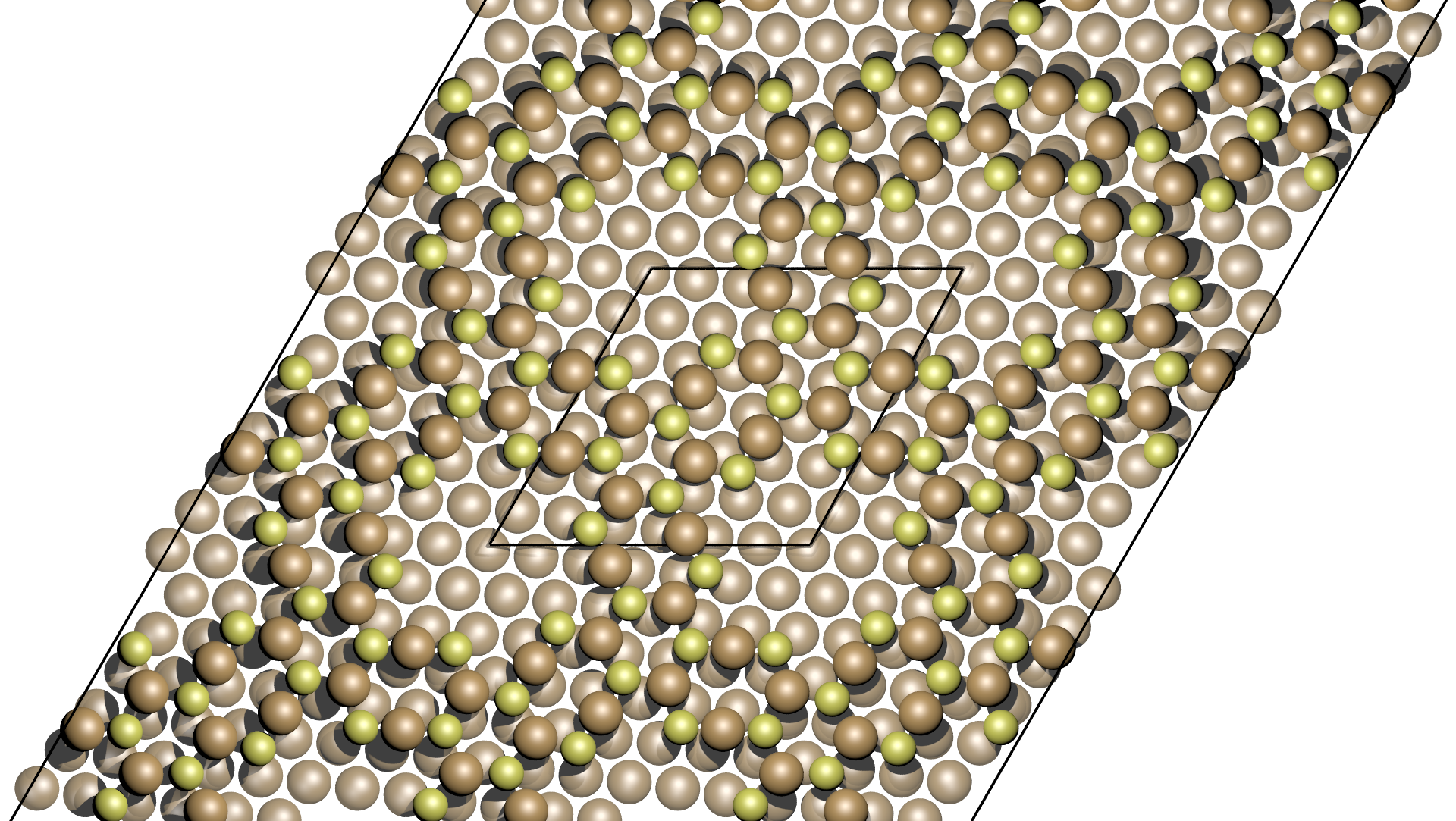}
\caption{Final sulfur induced $(\sqrt{43}\times \sqrt{43})$ surface reconstruction of Cu(111).}
\label{Fig4}
\end{center}
\end{figure}

The most stable structure for Cu$_{12}$S$_{12}$ stoichiometry presented in 
Fig.~\ref{Fig4} shows a $p6$ wallpaper group. It shows some similarities with
the (001) plane of the bulk covellite CuS which is a plane layer of 3 coordinated
Cu around three-fold rotation points of the cell.

\begin{figure}
\begin{center}
\includegraphics[width=0.5\textwidth]{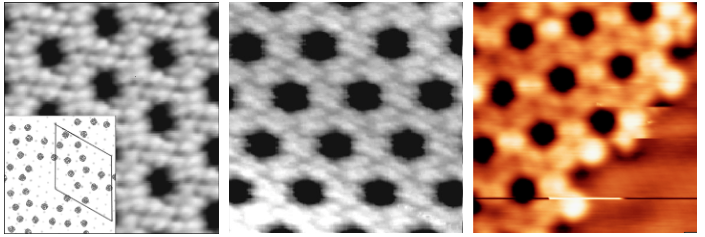}
\includegraphics[width=0.23\textwidth]{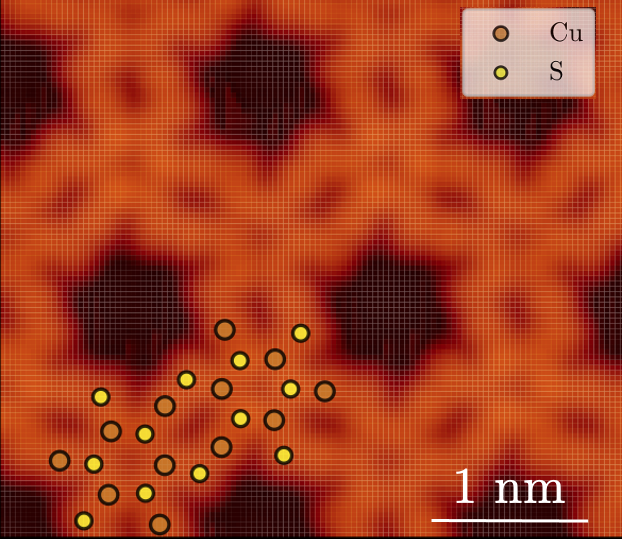}
\includegraphics[width=0.23\textwidth]{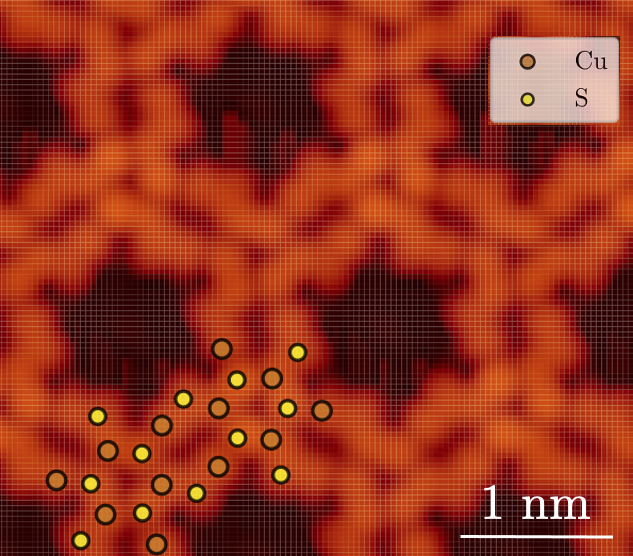}
\end{center}
\caption{(top) Experimental STM images of the $(\sqrt{43}\times\sqrt{43})$
Cu (111) surface reconstruction induced by sulfur.
Adapted with permission from E. Wahlström, I. Ekvall, H. Olin, S.-Å. Lindgren, and
L. Walldén Physical Review B \textbf{60}, 10699 (1999).~\cite{wahlstrom1999observation}
Copyright (1999) by the American Physical Society.
Adapted with permission from  E. Wahlström, I. Ekvall, T. Kihlgren, H. Olin, S.-Å. Lindgren, and
L. Walldén, Physical Review B \textbf{64}, 155406 (2001).~\cite{wahlstrom2001low}
Copyright (2001) by the American Physical Society.
Adapted with permission from D.-J. Liu, H. Walen, J. Oh, H. Lim, J. Evans, Y. Kim, and P. Thiel, The
Journal of Physical Chemistry C \textbf{118}, 29218 (2014).~\cite{liu2014search}
Copyright (2014) American Chemical Society.
(bottom) STM simulation of the energetically most stable 
Cu$_{12}$S$_{12}$ structure
for -0.7 V (left) and -0.2 V (right) biases.}
\label{Fig_stm}
\end{figure}

Experimental and simulated STM images are presented in Fig.~\ref{Fig_stm} at 
the top and bottom respectively. Simulated STM images were calculated at a constant current 
in the Tersoff-Hamann approximation for biases of -0.2 V and -0.7. They show 
excellent agreement with the experimental images, reproducing a honeycomb structure around
an empty center on the substrate. This structure is also consistent in terms of the
shortest S-S distances shown to be around 4 Å experimentally. We find them to be
3.95 Å for our Cu$_{12}$S$_{12}$ structure.

\section{Conclusion}


Recognizing the value of symmetry in searches, as a means to reduce the degrees of 
freedom of the search space, we have developed a way to incorporate structural 
candidates with imposed symmetry in the GOFEE method. The main contribution of 
the paper is the introduction of a so-called cascading symmetry rattle strategy. 
A key feature of this new methodology is the 
ability to dynamically alter the symmetry constraint of proposed structures during 
the global optimization procedure. This allows us to gradually explore from the most symmetric structures with the 
smallest configurational space, to those with lower symmetries and correspondingly much 
large configurational space. In doing so, we ensure proportional representation of 
structures with varying degrees of symmetry, despite the inherent bias towards 
non-symmetric $p1$ structures in the full configuration space. Additionally, the 
cascading symmetry strategy enables the use, and therefore advantages, of symmetry
constrained optimization without prior knowledge of the correct wallpaper group. 
It has been shown that for a number of systems, that
this strategy is much more efficient than not considering symmetry, in fact 
often comparable to considering only the correct wallpaper group, while still
allowing the algorithm to explore the full configurational space. 



These advances have enabled us to find a new structural model for the sulfur induced 
$(\sqrt{43}\times \sqrt{43})$ surface reconstruction of Cu(111). The 
efficiency of the method has allowed us to screen several stoichiometries, ultimately 
finding that a Cu$_{12}$S$_{12}$ reconstruction, that closely matches experimental 
observations, is the most energetically favorable.

\section{Supplementary Material}
In the supplementary information the coordinates of the Cu(111)-$\sqrt{43} \times \sqrt{43}$ structure are given together with a convergence test of its stability with respect to slab thickness.
\section{Acknowledgment}
This work has been supported by VILLUMFONDEN through InvestigatorGrant, Project No.16562, and by the Danish National Research Foundation through the Center of Excellence “InterCat” (GrantAgreement No.DNRF150).
\section{Code availability}
\label{sec:code_section}
Version 3.6.0 of the AGOX code is publicly available at \url{https://gitlab.com/agox/agox}, 
under a GNU GPLv3 license with documentation available at \url{https://agox.gitlab.io/agox}.

\section{Data availability}
The data that support the findings of this study are openly available at 
\url{https://gitlab.com/agox/agox_data}.

\bibliography{biblio}

\end{document}